\documentclass[conference,10pt]{IEEEtran}
\IEEEoverridecommandlockouts
 
\usepackage{amsmath}
\usepackage{cite}
\usepackage{amsmath,amssymb,amsfonts}
\usepackage{algorithm}
\usepackage{algpseudocode}
\usepackage{graphicx}
\usepackage{textcomp}
\usepackage{xcolor}
\usepackage{subcaption}
\usepackage{setspace}
\usepackage{booktabs}
\usepackage{array}
\usepackage{stfloats}
\usepackage{url}
\def\BibTeX{{\rm B\kern-.05em{\sc i\kern-.025em b}\kern-.08em
    T\kern-.1667em\lower.7ex\hbox{E}\kern-.125emX}}

\makeatletter

\newcommand{\Rmnum}[1]{\expandafter\@slowromancap\romannumeral #1@}
\makeatother

\makeatletter
\long\def\@makecaption#1#2{\ifx\@captype\@IEEEtablestring%
\footnotesize\begin{center}{\normalfont\footnotesize #1}\\
{\normalfont\footnotesize\scshape #2}\end{center}%
\@IEEEtablecaptionsepspace
\else
\@IEEEfigurecaptionsepspace
\setbox\@tempboxa\hbox{\normalfont\footnotesize {#1.}~~ #2}%
\ifdim \wd\@tempboxa >\hsize%
\setbox\@tempboxa\hbox{\normalfont\footnotesize {#1.}~~ }%
\parbox[t]{\hsize}{\normalfont\footnotesize \noindent\unhbox\@tempboxa#2}%
\else
\hbox to\hsize{\normalfont\footnotesize\hfil\box\@tempboxa\hfil}\fi\fi}
\makeatother

\begin{document}
\title{Conditional Diffusion-Based Point Cloud Imaging for UAV Position and Attitude Sensing \\
}
\author{
\IEEEauthorblockN{Xinhong~Dai\IEEEauthorrefmark{1},  Yuan Gao\IEEEauthorrefmark{1}, Hao Jiang\IEEEauthorrefmark{2}, Xiaojun~Yuan\IEEEauthorrefmark{2}, and~Xin Wang\IEEEauthorrefmark{1}}

\IEEEauthorblockA{\IEEEauthorrefmark{1}The Key Lab. for Info. Sci. of Electromagnetic Waves, College of Future Info. Tech., Fudan Uni., Shanghai.} 

\IEEEauthorblockA{\IEEEauthorrefmark{2}The National Key Lab. on Wireless Commun., Uni. of Electronic Sci. and Tech. of China, Chengdu.} 

\IEEEauthorblockA{Emails:\texttt{ \{xhdai24@m.,y\_gao23@m.,xwang11@\}fudan.edu.cn; \{jh@std.,xjyuan@\}uestc.edu.cn}}
}

\maketitle
\begin{abstract}
This paper studies an unmanned aerial vehicle (UAV) position and attitude sensing problem, where a base station equipped with an antenna array transmits signals to a predetermined potential flight region of a flying UAV, and exploits the reflected echoes for wireless imaging. The UAV is represented by an electromagnetic point cloud in this region that contains its spatial information and electromagnetic properties (EPs), enabling the unified extraction of UAV position, attitude, and shape from the reconstructed point cloud. To accomplish this task, we develop a generative
UAV sensing approach. The position and signal-to-noise ratio embedding
are adopted to assist the UAV features extraction from the estimated sensing channel under the measurement noise and channel variations. Guided by the obtained features, a conditional diffusion model is utilized to generate the point cloud. The simulation results demonstrate that the reconstructed point clouds via the proposed approach present higher fidelity compared to the competing schemes, thereby enabling a more accurate capture of the UAV attitude and shape information, as well as a more precise position estimation.
\end{abstract}


\IEEEpeerreviewmaketitle

\section{Introduction}
With the rapid development of the low-altitude economy (LAE) and the mission diversification of unmanned aerial vehicles (UAVs), the evolving sensing requirements of LAE have further driven the shift toward joint sensing of both position and attitude \cite{dai2025attitude}, as well as the classification of aerial targets \cite{Song2025cellular}. In this context, wireless sensing based on the base station (BS) is particularly attractive, owing to the strong air-to-ground line-of-sight (LoS) links and wider three-dimensional (3D) sensing coverage provided by antenna arrays. However, most existing studies mainly focus on the single attribute sensing of the UAV, such as position\cite{fang2023jtea,wen2023fast}, and thus cannot meet the growing demand for multidimensional UAV sensing.

From the perspective of electromagnetic propagation \cite{jiang2024paradigm}, characterizing or reconstructing these attributes from echo signals essentially forms a complex inverse problem, which maps echo signals from the UAV to an ``electromagnetic image''. Notably, endowed with robust capabilities in learning latent data distributions, generative artificial intelligence (GAI) has emerged as a promising paradigm for devising effective solutions to such an inverse problem\cite{gilpin2024generative}. For example, the study \cite{luo2021diffusion} presented a GAI-based point cloud generation method for 3D targets, where the generative model was used to transform the noise distribution to the distribution of a desired shape. Also, several studies \cite{wang2024generative,jiang2024electromagnetic} have employed some representative GAI models, e.g., diffusion models, for wireless sensing, including the human pose sensing and 3D target imaging. However, direct application of existing GAI-based methods to wireless imaging of highly dynamic UAVs faces several challenges: The extensive spatial range spanned by UAV mobility makes sensing and imaging across the entire flight domain computationally prohibitive, and hinders high-fidelity representation of UAV features. Also, the measurement noise and rapid channel variations make accurate extraction of intrinsic UAV features from echo signals more difficult.

Motivated by the above considerations, we investigate an array-based UAV position and attitude sensing problem. Specifically, we assume that the potential flight region of a flying UAV has been obtained in advance using some coarse position estimation methods before each sensing process. Then, a BS equipped with an antenna array transmits signals to this potential region and receives the corresponding echoes, based on which wireless imaging for the UAV is performed within this region. To visualize the sensed UAV, we adopt the electromagnetic point cloud to represent its spatial information and electromagnetic properties (EPs). We develop a generative UAV sensing approach to reconstruct the UAV point cloud. The position and signal-to-noise ratio (SNR) embedding are incorporated to assist the UAV features extraction from the estimated sensing channel under the measurement noise and channel variations. Guided by the obtained features, a conditional diffusion model is utilized to generate the point cloud. Numerical results show that the proposed approach achieves higher-quality point cloud imaging compared to the alternatives, resulting in the accurate UAV attitude and shape characterization, as well as the precise position estimation.


\section{System Model And Problem Formulation}
\begin{figure}[t]
    \centering
    \includegraphics[width=.75\linewidth]{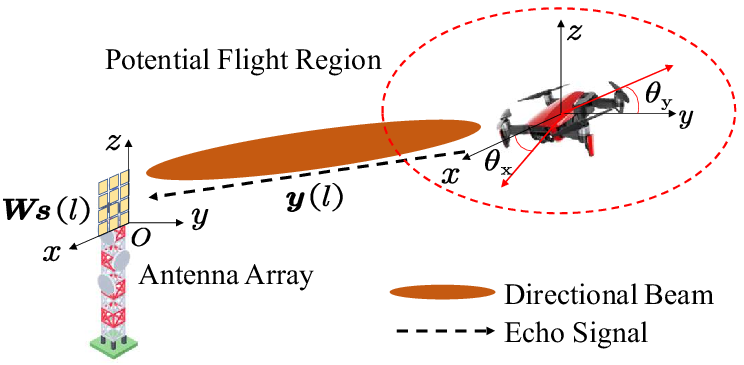}
    \caption{The considered UAV sensing system.}
    \label{system model}
    \vspace{-0.5 cm}
\end{figure}

As illustrated in Fig. \ref{system model}, we consider a UAV wireless sensing system, in which the BS transmits signals to a UAV flying in a 3D space and exploits the reflected echoes for UAV sensing. In this system, the BS is equipped with an antenna array consisting of $N_b = N_{\mathsf{x}} \times N_{\mathsf{z}}$ antennas for transmission and reception. All antennas are uniformly spaced at half of the carrier wavelength $\lambda_{\rm c} = \frac{c}{f_c}$, where $c$ denotes the speed of light and $f_c$ is the carrier frequency. In this paper, we only consider the LoS link between the UAV and the BS for analytical tractability. Before each sensing process, we assume that the BS has determined the potential flight region of the UAV via some coarse position estimation methods. Then, a rough beamforming strategy with a uniform angular interval is adopted based on a discrete Fourier transform (DFT) codebook. For each sensing process, we consider that the BS transmits $L$ symbols, and the $l$-th transmitted signal in vector form is given by:
\begin{align}
    \boldsymbol{x}(l) = \boldsymbol{W}\boldsymbol{s}(l)\in \mathbb{C}^{N_b\times 1},\quad l\in \mathcal{I}_L,
    \label{signal x}
\end{align}
where $\boldsymbol{W}=[\boldsymbol{w}_{1,1},\cdots,\boldsymbol{w}_{\imath,\jmath},\cdots,\boldsymbol{w}_{N_\mathsf{x},N_\mathsf{z}}]\in \mathbb{C}^{N_b \times N_b}$,$\imath\in \mathcal{I}_{N_\mathsf{x}}, \jmath\in \mathcal{I}_{N_\mathsf{z}}$, and $\boldsymbol{s}(l)\in \mathbb{C}^{N_b \times 1}$ denotes the $l$-th downlink symbol vector intended to the UAV. Here, the transmitted symbols are modeled as pseudo-random, satisfying $\mathbb{E}[\boldsymbol{s}(l)] = \mathbf{0}$ and $\mathbb{E}[\boldsymbol{s}(l)(\boldsymbol{s}(l))^\mathsf{H}] = \mathbf{I}_{N_b}$. 
\subsection{Sensing Model Based on Electromagnetic Scattering}
We assume that the antenna array operates in full-duplex mode, and the received signals at the BS can be modeled as 
\begin{align}
    \boldsymbol{Y} = \boldsymbol{H}\boldsymbol{X}+\boldsymbol{N}.
    \label{received echo signal}
\end{align}
In \eqref{received echo signal}, $\boldsymbol{Y}=[\boldsymbol{y}(1),\cdots,\boldsymbol{y}(L)]$ is the received signal matrix, and $\boldsymbol{N}=[\boldsymbol{n}(1),\cdots,\boldsymbol{n}(L)]$ represents the CSCG noise with the variance $\sigma^2$ at the BS, i.e., $\boldsymbol{n}(l)\sim \mathcal{CN}(\boldsymbol{0},\sigma^2 \mathbf{I}_{N_b})$. $\boldsymbol{H}\in \mathbb{C}^{N_b \times N_b}$ denotes the sensing channel matrix corresponding to the downlink transmission and signal reflection processes. We next detail the sensing model of the considered system based on the electromagnetic scattering modeling. 
Note that the EPs of the UAV are commonly manifested as spatial distributions of relative permittivity and conductivity, which capture the EP contrast between the UAV and the surrounding air medium. Referring to \cite{jiang2024electromagnetic}, assuming that scatter points on the UAV share the same EPs, the distributions can be formulated via the contrast function defined as
\begin{align}
     \label{contrast function}
    \chi(\boldsymbol{p}) = (\varepsilon(\boldsymbol{p})-1)+\frac{\text{j}\varrho(\boldsymbol{p})}{2\pi f_c\varepsilon_0},\quad \boldsymbol{p}\in \mathbb{R}^{3\times 1},
\end{align}
where $\varepsilon(\boldsymbol{p})$ and $\varrho(\boldsymbol{p})$ represent the relative permittivity and conductivity at any position $\boldsymbol{p}$ on the UAV, respectively. $\varepsilon_0$ is the vacuum permittivity. Then, according to \cite{bassen2003electric}, for the $l$-th transmitted signal, the incident electric field $\boldsymbol{E}_{\mathsf{i}}(\boldsymbol{p},l)$, at any position $\boldsymbol{p}$ in 3D space, can be interpreted as a linear superposition of the electric fields generated by all antennas:
\begin{align}
    \boldsymbol{E}_{\mathsf{i}}(\boldsymbol{p},l)=[\boldsymbol{m}_{1,1}(\boldsymbol{p}),\cdots,\boldsymbol{m}_{N_{\mathsf{x}},N_{\mathsf{z}}}(\boldsymbol{p})]\boldsymbol{x}(l),
    \label{Ei}
\end{align}
where $\boldsymbol{m}_{\imath,\jmath}(\boldsymbol{p}) = \int \overline{\overline{\boldsymbol{G}}}(\boldsymbol{p},\boldsymbol{p}'_{\imath,\jmath})\overline{\boldsymbol{J}}(\boldsymbol{p}'_{\imath,\jmath})\text{d}\boldsymbol{p}'_{\imath,\jmath}$ denotes the electric field at $\boldsymbol{p}$ generated by the $(\imath,\jmath)$-th antenna. Here,  $\overline{\boldsymbol{J}}(\boldsymbol{p}'{\imath,\jmath})$ denotes the equivalent current distribution induced at point $\boldsymbol{p}'{\imath,\jmath}$ on the $(\imath,\jmath)$-th antenna under unit current excitation; $\overline{\overline{\boldsymbol{G}}}(\boldsymbol{p},\boldsymbol{p}'_{\imath,\jmath})\in \mathbb{C}^{3\times 3}$ is the dyadic electric field Green’s function\cite{li1994electromagnetic}. Specifically, given any two points $\boldsymbol{p}_1$ and $\boldsymbol{p}_2$, $\overline{\overline{\boldsymbol{G}}}(\boldsymbol{p}_1, \boldsymbol{p}_2)$ is formulated as 
\begin{align}
\label{green funcion}
\overline{\overline{\boldsymbol{G}}}
(\boldsymbol{p}_1, \boldsymbol{p}_2)=[(g(r_{12})\overline{\boldsymbol{r}}_{12}\overline{\boldsymbol{r}}^{\mathsf{T}}_{12}-h(r_{12})\mathbf{I}_3]\frac{e^{\text{j} k_c r_{12}}}{4\pi r_{12}},
\end{align}
where $k_c = \frac{2\pi}{\lambda_c}$ is the wave number in the air, $r_{12}=\|\boldsymbol{p}_1 - \boldsymbol{p}_2\|$, $\overline{\boldsymbol{r}}_{12}=\frac{\boldsymbol{p}_1 - \boldsymbol{p}_2}{\|\boldsymbol{p}_1 - \boldsymbol{p}_2\|}$, $g(r_{12})=\frac{3}{k^2_c r^2_{12}}-\frac{3\text{j}}{k_c r_{12}}-1$, and $h(r_{12})=\frac{1}{k^2_c r^2_{12}}-\frac{\text{j}}{k_c r_{12}}-1$. When the incident electric field $\boldsymbol{E}_{\mathsf{i}}(\boldsymbol{p}, l)$ illuminates the UAV, according to the Lippmann-Schwinger equation\cite{chen2018computational}, the total electric field $\boldsymbol{E}_{\mathsf{tot}}(\boldsymbol{p},l)$ within the potential region of the UAV (denoted as $\mathcal{R}_{\mathsf{u}}$) is given by 
\begin{align}
   \boldsymbol{E}_{\mathsf{tot}}(\boldsymbol{p},l) = k^2_c\int_{\mathcal{R}_{\mathsf{u}}}\chi(\boldsymbol{p}')\overline{\overline{\boldsymbol{G}}}(\boldsymbol{p},\boldsymbol{p}')\boldsymbol{E}_{\mathsf{tot}}(\boldsymbol{p}',l)\text{d}\boldsymbol{p}'+\boldsymbol{E}_{\mathsf{i}}(\boldsymbol{p},l),\nonumber
\end{align}
where $\boldsymbol{p}'
\in\mathcal{R}_{\mathsf{u}}$. Then, the scattered electric field $\boldsymbol{E}_{\mathsf{s}}(\boldsymbol{p}_n,l)$ back to the $(\imath,\jmath)$-th antenna is similarly given by 
\begin{align}
    \label{wave eq Es}
    \boldsymbol{E}_{\mathsf{s}}(\boldsymbol{p}_{\imath,\jmath},l) = &k^2_c\int_{\mathcal{R}_{\mathsf{u}}}\chi(\boldsymbol{p}')\overline{\overline{\boldsymbol{G}}}(\boldsymbol{p}_{\imath,\jmath},\boldsymbol{p}')\boldsymbol{E}_{\mathsf{tot}}(\boldsymbol{p}',l)\text{d}\boldsymbol{p}'.
\end{align}
As an inverse process of \eqref{Ei}, $\boldsymbol{E}_{\mathsf{s}}(\boldsymbol{p}_n,l)$ will be converted into the electrical signal at each antenna. Thus, the $l$-th received signal at the BS is given by 
\begin{align}
    \label{signal yl}
    \boldsymbol{y}(l) = [\boldsymbol{E}_{\mathsf{s}}(\boldsymbol{p}_{1,1},l),\cdots,\boldsymbol{E}_{\mathsf{s}}(\boldsymbol{p}_{N_{\mathsf{x}},N_{\mathsf{z}}},l)]^\mathsf{T}\overline{\boldsymbol{v}}_p+\boldsymbol{n}(l),
\end{align}
where the unit vector $\overline{\boldsymbol{v}}_p\in \mathbb{R}^{3\times 1}$ denotes the polarization vector of the receiving array. Clearly, the equations from \eqref{Ei} to \eqref{signal yl} unfold the electromagnetic propagation mechanism in \eqref{received echo signal}. In other words, the EP distribution of the UAV in the 3D space is implicitly embedded in the received signals, as well as the sensing channel matrix, which can be utilized as prior knowledge to reconstruct the distribution of $\chi(\boldsymbol{p})$. To this end, we adopt the least squares (LS) method to estimate the sensing channel, denoted as $\boldsymbol{H}_{\mathsf{est}}$.

\subsection{Problem Formulation}
In this paper, to effectively estimate the position and attitude of the UAV within the potential region, a point cloud-based representation is adopted to visualize the EP distribution of the UAV. As mentioned before, we denote the center of the predetermined region before the sensing by $\boldsymbol{q}_{\mathsf{pre}} = [\tilde{q}_x,\tilde{q}_y,\tilde{q}_z]^{\mathsf{T}}$.
Then, we represent the UAV as a 5D point cloud, which is composed of $M$ normalized points, i.e., 
$\mathcal{P} = \{\boldsymbol{p}_i |i\in \mathcal{I}_M\}$. Each 5D point in $\mathcal{P}$ contains two parts, including the 3D coordinate information and 2D EPs information: 
\begin{align}
    \label{5D pc}
    \boldsymbol{p}_i = \left[\frac{x_i-\tilde{q}_x}{s_x},\frac{y_i-\tilde{q}_y}{s_y},\frac{z_i-\tilde{q}_z}{s_z},\frac{\varepsilon}{\varepsilon_0},\frac{\varrho}{2\pi f_c\varepsilon_0}\right]^{\mathsf{T}},
\end{align}
where $x_i$, $y_i$, and $z_i$ represent the coordinates of the $i$-th point along the corresponding dimension; $s_x$, $s_y$, and $s_z$ denote the standard deviations of the reconstruction region, which hinge on the UAV size and the error threshold of the coarse position estimation. Note that $\mathcal{P}$ provides a discrete representation of the intrinsic properties of the UAV and can be regarded as a collection of samples drawn from an underlying high-dimensional distribution $p(\mathcal{P})$. Therefore, the UAV position and attitude sensing problem (i.e., a point cloud reconstruction problem) can be formulated as the maximization of the conditional probability given the measurements $\boldsymbol{H}_\mathsf{est}$, i.e., 
\begin{align}
     \label{problem formulation 1}
    \arg \max \quad p(\mathcal{P}|\boldsymbol{H}_\mathsf{est}).
\end{align}
In general, this problem is difficult to solve because it remains challenging to explicitly characterize both $p(\mathcal{P})$ and $p(\mathcal{P}|\boldsymbol{H}_\mathsf{est})$, owing to the intricate mapping from the multidimensional information of the UAV to the channel matrix. To address this, we next leverage the powerful data distribution learning capabilities of generative learning and propose an approach by following the GAI principle.
\vspace{-0.1 cm}
\section{Proposed Approach}
\begin{figure}[t]
    \centering
\includegraphics[width=.9\linewidth]{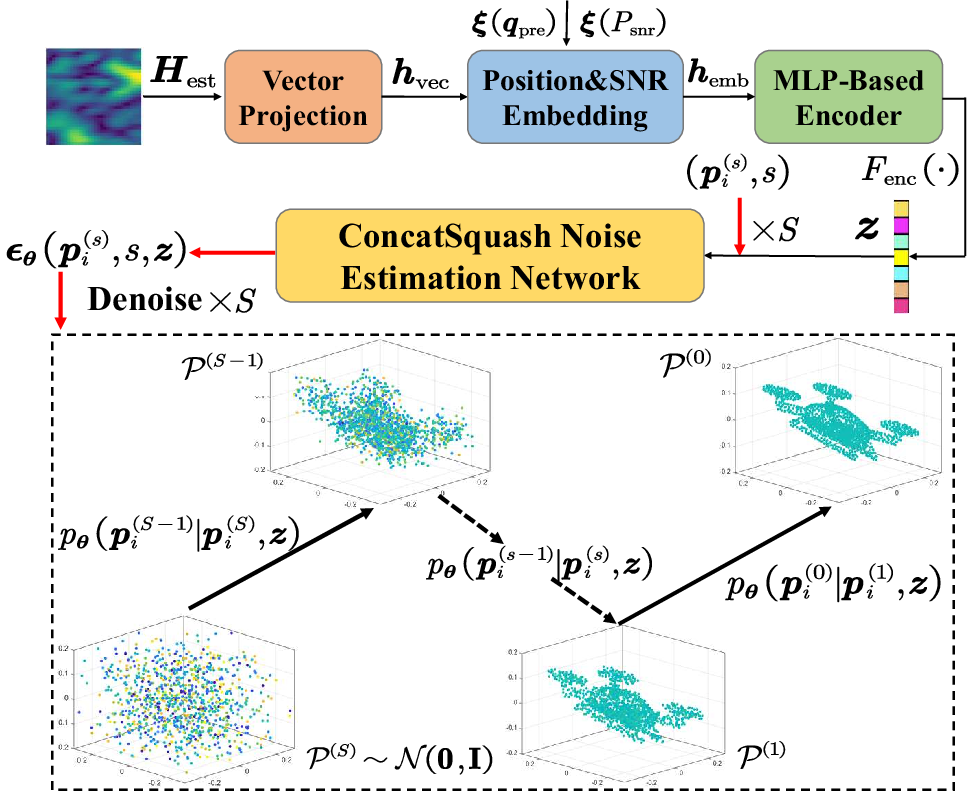}
    \caption{The illustration of the proposed approach.}
    \label{AUGUST}
    \vspace{-0.5 cm}
\end{figure}

From a generative learning perspective, a direct method to solve problem \eqref{problem formulation 1} is to train a parameterized probabilistic model $p_{\boldsymbol{\theta}}(\mathcal{P}|\boldsymbol{H}_\mathsf{est})$ to approximate $p(\mathcal{P}|\boldsymbol{H}_\mathsf{est})$. Hence, the training objective is to maximize the log-likelihood $\log p_{\boldsymbol{\theta}}(\mathcal{P}|\boldsymbol{H}_\mathsf{est})$. Nevertheless, direct maximization of $\log p_{\boldsymbol{\theta}}(\mathcal{P}|\boldsymbol{H}_\mathsf{est})$ is generally intractable due to latent yet inevitable physical factors, such as limited bandwidth and measurement noise. These may result in the weak and ambiguous representations of $\boldsymbol{H}_\mathsf{est}$ for UAV EPs. To tackle this, we next introduce a latent variable $\boldsymbol{z}$ to explicitly encode intrinsic but partially unobservable features from $\boldsymbol{H}_\mathsf{est}$, and reconstruct the point cloud via the expected inference pathway $\boldsymbol{H}_\mathsf{est}\to\boldsymbol{z}\to \mathcal{P}$, as shown in Fig. \ref{AUGUST}.
For a flying UAV, the sensing channel $\boldsymbol{H}_{\mathsf{est}}$ varies not only with changes in the UAV flight status but also depends strongly on the UAV’s spatial location and beamforming design. In particular, the phase associated with each scattering point follows an exponential phase mapping with respect to its spatial coordinates. Thus, to facilitate stable UAV feature extraction by the encoder under continuously varying UAV positions and channel conditions, we introduce the positional embedding and SNR embedding in the channel encoder.

Specifically, we first vectorize the obtained $\boldsymbol{H}_\mathsf{est}$ and project it linearly into a $d_p$-dimensional channel vector based on a fully connected layer with parameters $\boldsymbol{W}_\text{vec}$ and $\boldsymbol{b}_p$, i.e.,
\begin{align}
    \label{h vec}
    \boldsymbol{h}_\text{vec} = \boldsymbol{W}_p\text{vec}(\boldsymbol{H}_\mathsf{est}) +\boldsymbol{b}_p.
\end{align}
Given the potential region center and the SNR (denoted as $P_{\text{snr}} = \frac{\|\boldsymbol{H}\boldsymbol{X}\|^2_F}{\|\boldsymbol{N}\|^2_F}$), the position encoding function $\boldsymbol{\xi}(\boldsymbol{q}_{\mathsf{pre}})$ and the SNR encoding function $\boldsymbol{\xi}(P_{\text{snr}})$, defined as two higher-dimensional Fourier feature space mappings $\mathbb{R}^{3\times(2 d_\xi+1)}$, are employed to address the issue of capturing high-frequency variations in the sensing channel, i.e.,
\begin{align}
    \label{position encoding}
    \boldsymbol{\xi}(\boldsymbol{q}_{\mathsf{pre}}) = &[\boldsymbol{q}_{\mathsf{pre}};\sin(2^0\pi \boldsymbol{q}_{\mathsf{pre}});\cos(2^0\pi \boldsymbol{q}_{\mathsf{pre}});\cdots;\nonumber\\
    &\sin(2^{d_\xi-1}\pi \boldsymbol{q}_{\mathsf{pre}});\cos(2^{d_\xi-1}\pi \boldsymbol{q}_{\mathsf{pre}})],
\end{align}
and $\boldsymbol{\xi}(P_{\text{snr}})\in\mathbb{R}^{2 d_\xi+1}$ has a similar form by replacing $\boldsymbol{q}_{\mathsf{pre}}$ with $P_{\text{snr}}$. $d_\xi$ is the number of encoding frequencies. After this transform, high-frequency spatial and value variations become two linear combinations of fixed harmonic bases. Besides, since wireless channels are continuous physical fields whose statistics are strongly coupled to the UAV position, the conventional additive positional embeddings are thus not very applicable to the considered problem. Motivated by \cite{xing2025multi}, we adopt a multiplicative strategy for positional embedding, and the embedded channel vector input to the encoder is given by
\begin{align}
    \label{h emb}
    \boldsymbol{h}_\text{emb} = [\boldsymbol{\omega}_1(\boldsymbol{\xi}(\boldsymbol{q}_{\mathsf{pre}}))\odot \boldsymbol{h}_\text{vec};\boldsymbol{\omega}_2(\boldsymbol{\xi}(P_{\text{snr}}))],
\end{align}
where $\boldsymbol{\omega}_1(\cdot)$ and $\boldsymbol{\omega}_2(\cdot)$ are two shallow fully connected heads that explicitly capture the position- and SNR-dependent effects, while $\boldsymbol{h}_\text{vec}$ focuses on target-related information that is comparatively stable across varying UAV positions and channel conditions. Given $\boldsymbol{h}_\text{emb}$, we proceed to extract a feature vector $\boldsymbol{z}$ with dimension $d_z$ from this embedded channel vector via an MLP-based feature encoder, i.e., $\boldsymbol{z} = F_{\text{enc}}(\boldsymbol{h}_\text{emb})$. The extracted feature vector $\boldsymbol{z}$ encodes the intrinsic properties of the UAV, which can be used as a conditional input for the following generative reconstruction module.

Based on the obtained high-dimensional features $\boldsymbol{z}$, we aim to design the decoder $p_{\boldsymbol{\theta}}(\mathcal{P}|\boldsymbol{z})$ by employing the conditional diffusion model \cite{luo2021diffusion}, as detailed in the following. 

\textit{1) Forward Diffusion Process}: To clearly present the diffusion and denoising processes, $\mathcal{P}^{(1)}, \ldots, \mathcal{P}^{(S)}$ are denoted as a sequence of point clouds sequentially generated from the original point cloud $\mathcal{P}^{(0)} = \{\boldsymbol{p}^{(0)}_i\}^{M}_{i=1}$ defined in \eqref{5D pc}, up to the maximum time step $S$ (ech step is indexed by $s$). Since the 5D point cloud $\mathcal{P}^{(0)} = \{\boldsymbol{p}^{(0)}_i\}^{M}_{i=1}$ reflects the holistic distribution of UAV EPs, $p_{\boldsymbol{\theta}}(\mathcal{P}^{(0)}|\boldsymbol{z})$ can be factorized into $M$ independent samples drawn from the point distribution $q(\boldsymbol{p}^{(0)}_i|\boldsymbol{z})$. In the forward diffusion process, the transition from the original distribution $q(\boldsymbol{p}^{(0)}_i)$ to a noise-alike one $q(\boldsymbol{p}^{(S)}_i)$ can be modeled as a Markov chain:
\begin{align}
    \label{forward markov chain}
 q(\boldsymbol{p}^{(1:S)}_i&|\boldsymbol{p}^{(0)}_i)  = {\prod\nolimits_{s=1}^{S}}  q(\boldsymbol{p}^{(s)}_i|\boldsymbol{p}^{(s-1)}_i)\nonumber\\
    &={\prod\nolimits_{s=1}^{S}}\mathcal{N}(\boldsymbol{p}^{(s)}_i;\sqrt{1-\beta_s}\boldsymbol{p}^{(s-1)}_i,\beta_s\mathbf{I}).
\end{align}
Here, $\beta_s$ is a hyperparameter controlling the noise intensity and increases linearly with the step index $s$. Then, according to Bayes’ rule and using the renormalization technique, the calculation of $\boldsymbol{p}^{(s)}_i$ at an
arbitrary time step $s$ can be simplified to
\begin{align}
    \label{forward q s}    q(\boldsymbol{p}^{(s)}_i|\boldsymbol{p}^{(0)}_i)  = \mathcal{N}(\boldsymbol{p}^{(s)}_i;\sqrt{\overline{\alpha}_s}\boldsymbol{p}^{(0)}_i,\sqrt{1-\overline{\alpha}_s}\mathbf{I}),
\end{align}
where $\alpha_s = 1-\beta_s$ and $\overline{\alpha}_s = \prod^{s}_{s'=1}\alpha_{s'} = {\alpha}_s\overline{\alpha}_{s-1}$.

\textit{2) Reverse Diffusion Process}: Under the diffusion framework, the reconstruction process corresponds to the reverse of the forward diffusion process. From \eqref{forward markov chain} and \eqref{forward q s}, the reverse transition probability conditioned on $\boldsymbol{p}^{(0)}_i$ is given by
\begin{align}
    \label{reverse q s}
    q(\boldsymbol{p}^{(s-1)}_i|\boldsymbol{p}^{(s)}_i,\boldsymbol{p}^{(0)}_i)  = \mathcal{N}(\boldsymbol{p}^{(s-1)}_i;\tilde{\boldsymbol{\mu}}_s (\boldsymbol{p}^{(s)}_i,\boldsymbol{p}^{(0)}_i),\tilde{\beta}_s\mathbf{I}),
\end{align}
with 
\begin{align}
    \label{mu s}
\tilde{\boldsymbol{\mu}}_s (\boldsymbol{p}^{(s)}_i,\boldsymbol{p}^{(0)}_i) & = \frac{\sqrt{\overline{\alpha}_{s-1}}\beta_s}{1-\overline{\alpha}_{s}}\boldsymbol{p}^{(0)}_i+\frac{\sqrt{\alpha_s}(1-\overline{\alpha}_{s-1})}{1-\overline{\alpha}_{s}}\boldsymbol{p}^{(s)}_i\nonumber\\
& = \frac{1}{\sqrt{\alpha_s}}\left(\boldsymbol{p}^{(s)}_i-\tilde{\alpha}_s\boldsymbol{\epsilon}_i^{(s)}\right).
\end{align}
Here, $\tilde{\beta}_s = \frac{1-\overline{\alpha}_{s-1}}{1-\overline{\alpha}_{s}}\beta_s$, $\tilde{\alpha}_s=\frac{1-\alpha_s}{\sqrt{1-\overline{\alpha}_{s}}}$, and $\boldsymbol{\epsilon}_i^{(s)} = \frac{1}{\sqrt{1-\overline{\alpha}_{s}}}(\boldsymbol{p}^{(s)}_i-\sqrt{\overline{\alpha}_{s}}\boldsymbol{p}^{(0)}_i)$. However, during reconstruction, the original point cloud $\mathcal{P}^{(0)}$ is unknown, and the only available conditioning information is $\boldsymbol{z}$ encoded from the sensing channel. To tackle this, the reverse process is implemented by introducing a sequence of learnable, parameterized transition posteriors conditioned on $\boldsymbol{z}$, i.e.,
\begin{align}
    \label{reverse markov chain}
    p_{\boldsymbol{\theta}}(\boldsymbol{p}^{(0:S)}_i|\boldsymbol{z}) = p(\boldsymbol{p}^{(S)}_i)\prod\nolimits_{s=1}^{S} p_{\boldsymbol{\theta}}(\boldsymbol{p}^{(s-1)}_i|\boldsymbol{p}^{(s)}_i,\boldsymbol{z}).
\end{align}
Here, $p(\boldsymbol{p}^{(S)}_i)$ is a predetermined Gaussian distribution from which the initial random point cloud $\mathcal{P}^{(S)}$ is sampled. $p_{\boldsymbol{\theta}}(\boldsymbol{p}^{(s-1)}_i|\boldsymbol{p}^{(s)}_i,\boldsymbol{z})$ denotes the learnable reverse transition probability that follows the Gaussian distribution:
\begin{align}
    \label{reverse p theta}
    p_{\boldsymbol{\theta}}(\boldsymbol{p}^{(s-1)}_i|\boldsymbol{p}^{(s)}_i,\boldsymbol{z}) = \mathcal{N}(\boldsymbol{p}^{(s-1)}_i;{\boldsymbol{\mu}}_{\boldsymbol{\theta}} (\boldsymbol{p}^{(s)}_i,s,\boldsymbol{z}),\tilde{\beta}_s\mathbf{I}).
\end{align}
Since \eqref{reverse p theta} is a parameterized approximation of \eqref{reverse q s}, ${\boldsymbol{\mu}}_{\boldsymbol{\theta}} (\boldsymbol{p}^{(s)}_i,s,\boldsymbol{z})$ has a similar form as $\tilde{\boldsymbol{\mu}}_s (\boldsymbol{p}^{(s)}_i,\boldsymbol{p}^{(0)}_i)$, which is given by
\begin{align}
    \label{mu theta}
{\boldsymbol{\mu}}_{\boldsymbol{\theta}} (\boldsymbol{p}^{(s)}_i,s,\boldsymbol{z}) = \frac{1}{\sqrt{\alpha_s}}\left(\boldsymbol{p}^{(s)}_i-\tilde{\alpha}_s\boldsymbol{\epsilon}_{\boldsymbol{\theta}}(\boldsymbol{p}^{(s)}_i,s,\boldsymbol{z})\right),
\end{align}
where $\boldsymbol{\epsilon}_{\boldsymbol{\theta}}(\boldsymbol{p}^{(s)}_i,s,\boldsymbol{z})$ denotes the learning-dependent noise estimator. To implement this noise predictor, we adopt a unified noise estimation network composed of a series of $L_{\text{cs}}$ ConcatSquash layers proposed in \cite{grathwohl2018ffjord}, and each layer can be formulated as:
\begin{align}
    \label{concatsquash}
    \boldsymbol{\eta}^{\ell+1} = (\boldsymbol{W}^{\ell}_1\boldsymbol{\eta}^{\ell}+\boldsymbol{b}^{\ell}_1)\odot\varsigma(\boldsymbol{W}^{\ell}_2\boldsymbol{c}+\boldsymbol{b}^{\ell}_2)+\boldsymbol{W}^{\ell}_3\boldsymbol{c}.
\end{align}
Here, $\boldsymbol{\eta}^{\ell}$ and $\boldsymbol{\eta}^{\ell+1}$ are the input and output of the $\ell$-th layer. $\varsigma(\cdot)$ is the Sigmoid function. $\boldsymbol{W}^{\ell}_1$, $\boldsymbol{W}^{\ell}_2$, $\boldsymbol{W}^{\ell}_3$, $\boldsymbol{b}^{\ell}_1$, and $\boldsymbol{b}^{\ell}_2$ denote the trainable weights and biases in the network. The context vector $\boldsymbol{c} = [\overline{s},\sin(\overline{s}),\cos({\overline{s}}),\sin(2\overline{s}),\cos(2\overline{s});(\boldsymbol{z})^\mathsf{T}]^\mathsf{T}$ incorporates the embedded features of the normalized diffusion step $\overline{s} = \frac{s}{S}$ and the latent code $\boldsymbol{z}$. The swish function is used as the activation function between two adjacent layers. The input of the whole network is the 5D point $\boldsymbol{p}^{(s)}_i$, and the output is the estimated noise $\boldsymbol{\epsilon}_{\boldsymbol{\theta}}(\boldsymbol{p}^{(s)}_i,s,\boldsymbol{z})$.

In the reverse diffusion process, the training goal is to maximize the log-likelihood $\log p_{\boldsymbol{\theta}}(\mathcal{P}^{(0)}|\boldsymbol{z})$ given the original UAV point cloud and the latent code $\boldsymbol{z}$. Clearly, the comparison between the parameterized reverse Markov chain and the original one implicitly encourages the learned reverse process to match the true posterior over the diffusion path. In other words, the diffusion model, guided by conditional features $\boldsymbol{z}$, gradually fits the noise and posterior transition probability, i.e, minimizing $D_{\text{KL}}(q(\boldsymbol{p}^{(s-1)}_i|\boldsymbol{p}^{(s)}_i,\boldsymbol{p}^{(0)}_i)\|p_{\boldsymbol{\theta}}(\boldsymbol{p}^{(s-1)}_i|\boldsymbol{p}^{(s)}_i,\boldsymbol{z}) )$ for each step $s$, thereby denoising and generating the original data distribution. Here, $D_{\text{KL}}(p(\cdot)\|q(\cdot))$ denotes the Kullback–Leibler (KL) divergence from distribution $p(\cdot)$ to distribution $q(\cdot)$. Hence, the loss function of the diffusion-based noise estimator $\boldsymbol{\epsilon}_{\boldsymbol{\theta}}(\boldsymbol{p}^{(s)}_i,s,\boldsymbol{z})$ can be given by
\begin{align}
    \label{loss concatquash}
    \mathcal{L}_{\text{DM};i}^{(s)} = C_{\alpha_s,\beta_s}\|\boldsymbol{\epsilon}_i^{(s)}-\boldsymbol{\epsilon}_{\boldsymbol{\theta}}(\boldsymbol{p}^{(s)}_i,s,\boldsymbol{z})\|^2,
\end{align}
where $C_{\alpha_s,\beta_s}$ is a learning-free parameter. Note that the numerical and spatial distribution complexities of the 3D positions and those for EPs in UAV point clouds exhibit significantly different distributions. To ensure the effectiveness of the reconstruction, denote the distribution of the training data by $p_{\text{train}}(\mathcal{P}^{(0:S)},\boldsymbol{H}_\mathsf{est})$, and we can train these two parts separately through a weighted training loss as
\begin{align}
    \label{loss joint}
    \mathcal{L}_{\text{train}} = \mathbb{E}_{p_{\text{train}},s,\boldsymbol{z},\boldsymbol{\epsilon}_i^{(s)}}\big[\gamma_{\text{pos}} \mathcal{L}_{\text{pos};i}^{(s)}+\gamma_{\text{EP}} \mathcal{L}_{\text{EP};i}^{(s)}\big],
\end{align}
where $\mathcal{L}_{\text{pos};i}^{(s)} = \|\boldsymbol{\epsilon}_{\text{pos};i}^{(s)}-\boldsymbol{\epsilon}_{\text{pos};\boldsymbol{\theta}}(\boldsymbol{p}^{(s)}_i,s,\boldsymbol{z})\|^2$ and $\mathcal{L}_{\text{EP};i}^{(s)} = \|\boldsymbol{\epsilon}_{\text{EP};i}^{(s)}-\boldsymbol{\epsilon}_{\text{EP};\boldsymbol{\theta}}(\boldsymbol{p}^{(s)}_i,s,\boldsymbol{z})\|^2$ denote the diffusion loss for 3D position and EPs of the $i$-th 5D point, respectively; Their dimensions correspond to the definition in \eqref{5D pc}. $\gamma_{\text{pos}/\text{EP}}$ is an adjustable weighted coefficient. For the inference, we can obtain the reconstructed UAV point cloud (denoted as $\hat{\mathcal{P}}^{(0)} = \{\hat{\boldsymbol{p}}^{(0)}_i\}^{M}_{i=1}$) from the initial noise samples by successively sampling $p_{\boldsymbol{\theta}}(\boldsymbol{p}^{(0:S)}_i|\boldsymbol{z})$ as shown in Fig. \ref{AUGUST}.

\section{Numerical Results}
Assume that the BS is centered at $(0,0,0)$ m, and the antenna array operates at $f_c = 3$ GHz. The noise power $\sigma^2=-120$ dBm and the transmission power is set to $P_{\mathsf{s}} \in [10,40]$ dBm. The antenna array is set to $N_{\mathsf{b}}=16\times2$ and the number of transmitted symbols $L=32$ by default. To construct the UAV point cloud samples, we select five typical shapes of rotary-wing UAVs from \cite{chang2015shapenet}. We perform surface discrete random sampling and local coordinate system transformation on all UAVs to obtain UAV point cloud samples with different attitudes and shapes. The number of points in each UAV point cloud is fixed at $M=1000$ for both training and inference, and the standard deviations of the reconstruction region are set to $s_{x/y/z}=0.85$ m. In addition, a total of 50,000 UAV point cloud samples (10,000 for each shape) are generated for training, validation, and testing with a ratio of 8:1:1. These samples are independently and randomly generated based 1000 random discrete trajectories and subject to the flight range $q_{\rm{x/y}}\in[10,30]$ m, $q_{\rm{z}}\in[10,20]$ m, and the tilt angle range $\theta_{\rm{x/y}}\in[-30^{\circ},30^{\circ}]$, $\theta_{\rm{z}}\in[-180^{\circ},180^{\circ}]$ as illustrated in Fig. \ref{system model}. Each trajectory is composed of 50 discrete points with a uniform time interval $\Delta_t = 0.2$ s, where the UAV velocity, acceleration, and corresponding tilt angles all conform to basic constraints detailed in \cite{dai2025attitude}. Moreover, the relative permittivity and conductivity of the UAVs are randomly set within the ranges $\varepsilon\in[1.5,5]$ and $\varrho\in[1,10]$ mS/m. The integral equations of the electromagnetic scattering process are tackled in discrete form based on methods of moments (MoM) \cite{jiang2024electromagnetic}.

\begin{figure*}[t]
\centering
\subcaptionbox{\footnotesize Ground Truth 1\label{GT1}}%
{\includegraphics[width=0.17\textwidth]{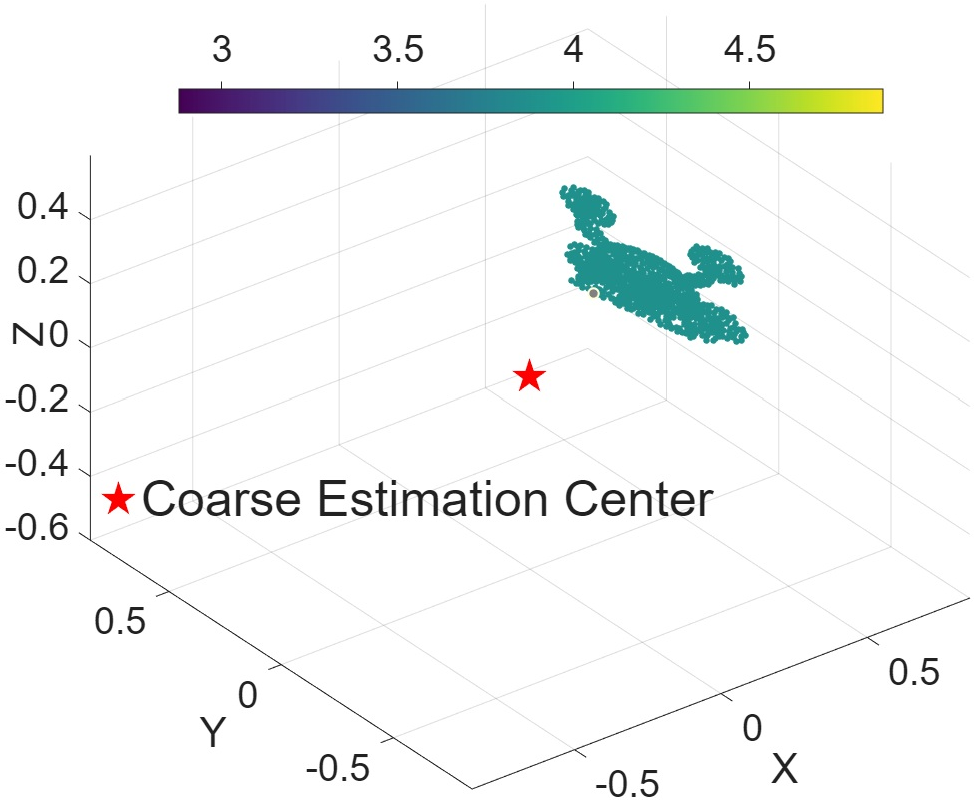}}
\subcaptionbox{\footnotesize Proposed \label{Proposed_Noncenter}}%
{\includegraphics[width=0.17\textwidth]{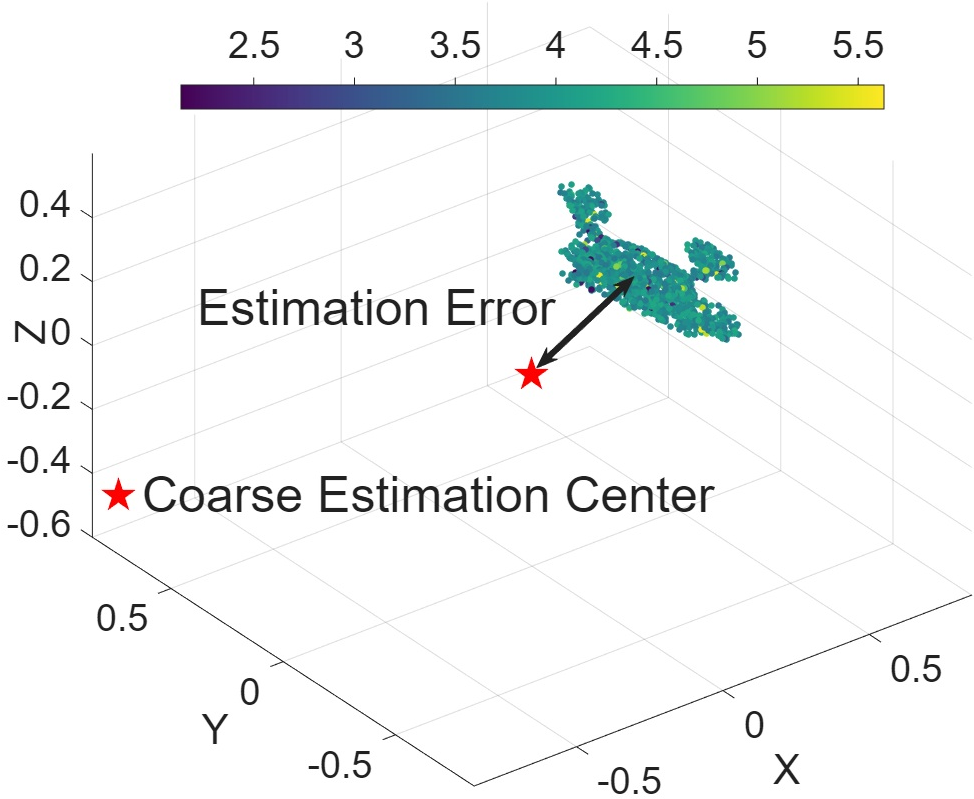}}
\subcaptionbox{\footnotesize Ground Truth 2\label{GT2}}%
{\includegraphics[width=0.155\textwidth]{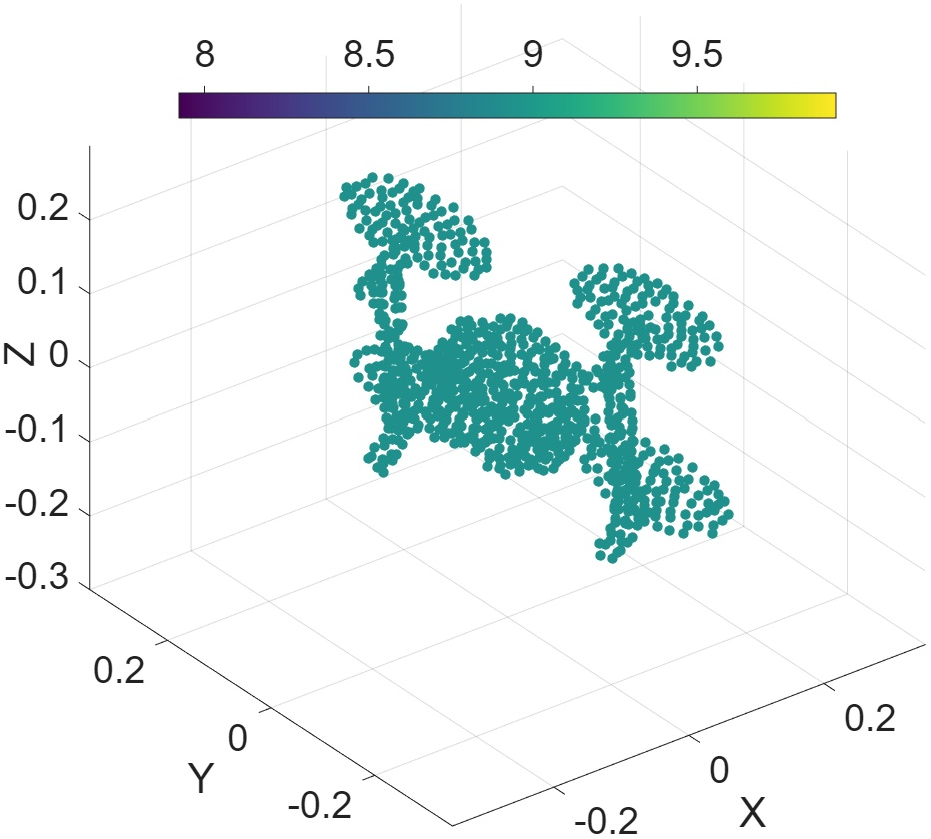}}
\subcaptionbox{\footnotesize Proposed\label{Proposed}}%
{\includegraphics[width=0.155\textwidth]{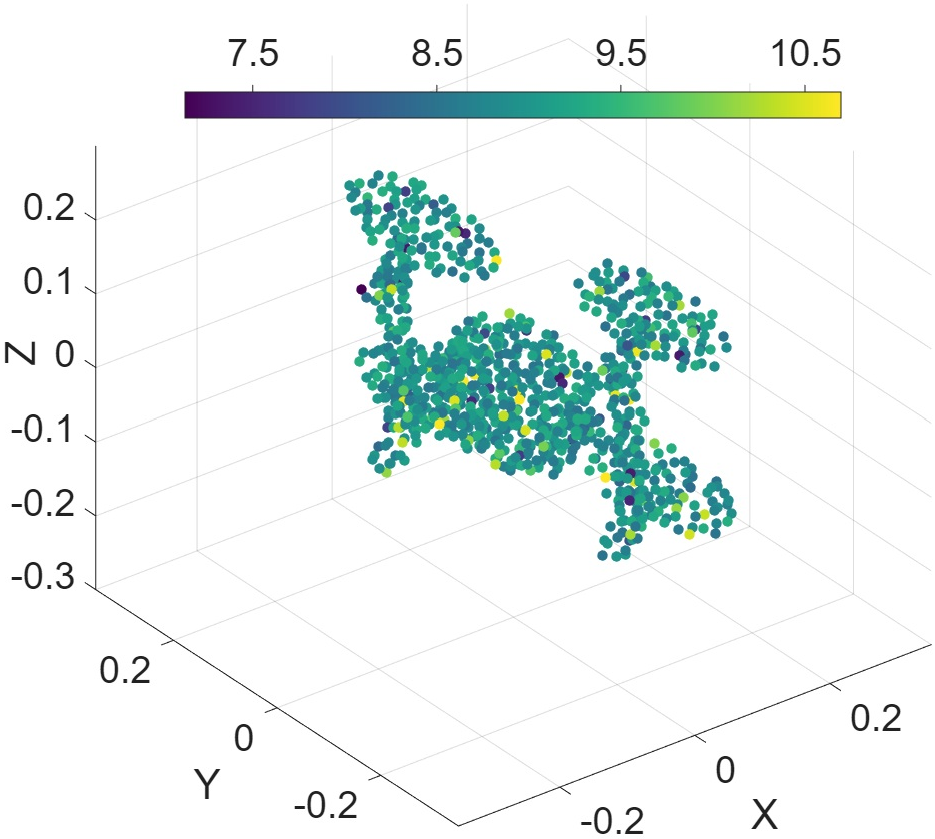}}
\subcaptionbox{\footnotesize NoEBD\label{NoEBD}}%
{\includegraphics[width=0.155\textwidth]{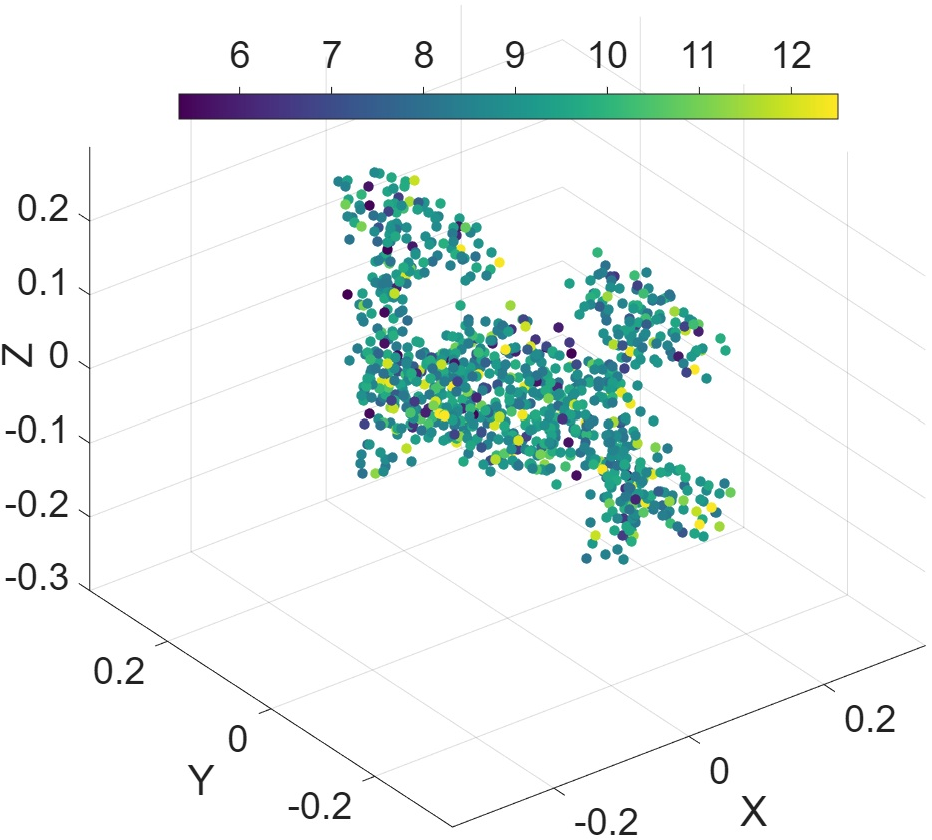}}
\subcaptionbox{\footnotesize Direc-Channel\label{Direct}}%
{\includegraphics[width=0.155\textwidth]{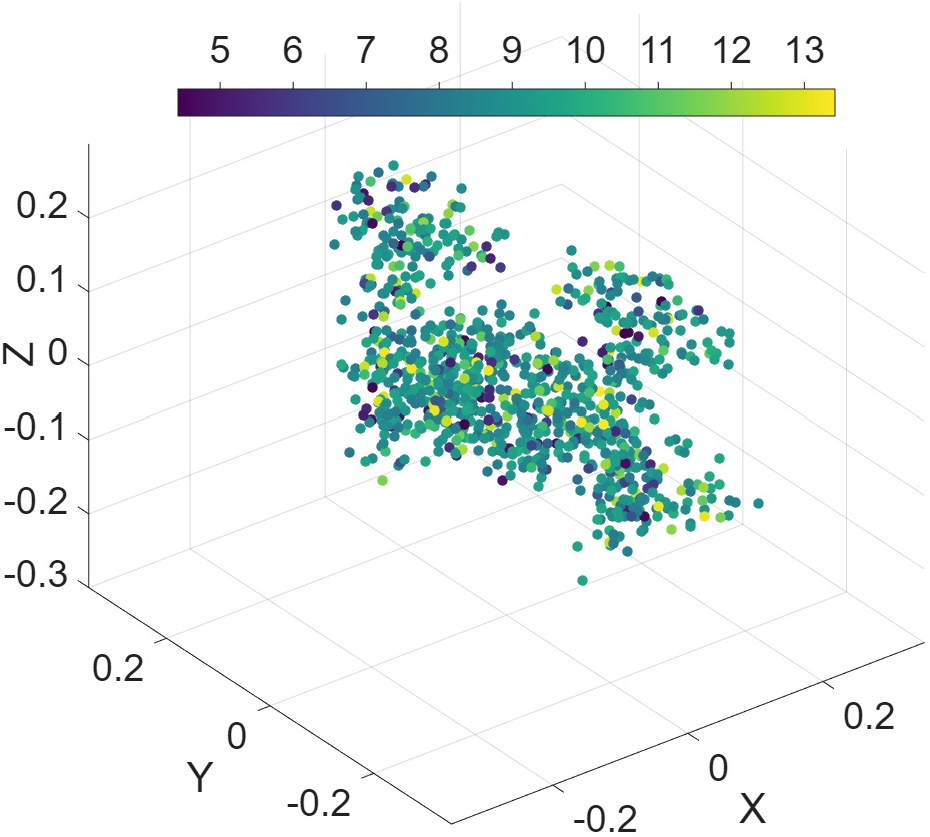}}
\caption{(a)-(b): Potential region view of the point cloud reconstruction result with the relative permittivity based on the proposed approach; (c)-(f): Centralized view of the point cloud reconstruction results with the conductivity based on different approaches; $\text{SNR}$ is fixed as 20 dB.}
\label{PC}
\vspace{-0.3 cm}
\end{figure*}
For the diffusion model, 12 ConcatSquash layers of the noise estimation network are structured as 5-16-64-128-256-512-1024-512-256-128-64-16-5. The noise intensity parameter $\beta_s$ linearly increases from 0.0001 to 0.05 over $S=200$ diffusion steps. Regarding the training configurations, the MLP consists of 6 layers, and the other fully-connected networks contain 2 layers. The batch size and training epochs are respectively set to 256 and 200, with a linearly decaying learning rate from $10^{-4}$ to $10^{-5}$. The Adam optimizer is used for optimization. Since we focus more on sensing for the UAV position, attitude, and shape, the weighted coefficients in are set as $\gamma_{\text{pos}}=0.9$ and $\gamma_{\text{EP}}=0.1$. The remaining training parameters are as follows: $d_v = 256$, $d_z = 512$, and $d_{\xi} =10$.

To evaluate the reconstruction performance of the proposed approach, we adopt an evaluation metric, named as weighted distance (WD), that combines the mean Chamfer distance and the MSE of the position estimation:
\begin{align}
    \label{WD-compute}
   & \text{WD}(\text{dB}) = 10\log_{10}\mathbb{E}\bigg[\frac{1}{M}\sum_{\hat{\boldsymbol{p}}\in\hat{\mathcal{P}}_c^{(0)}}\min_{\boldsymbol{p}\in\mathcal{P}_c^{(0)}}\|\hat{\boldsymbol{p}}-\boldsymbol{p}\|^2\nonumber\\
    &+ \frac{1}{M}\sum_{\boldsymbol{p}\in\mathcal{P}_c^{(0)}}\min_{\hat{\boldsymbol{p}}\in\hat{\mathcal{P}}_c^{(0)}}\|\boldsymbol{p}-\hat{\boldsymbol{p}}\|^2+\|\boldsymbol{q}-\boldsymbol{q}_{\mathsf{est}}\|^2\bigg],
\end{align}
where $\mathcal{P}_c^{(0)}$ and $\hat{\mathcal{P}}_c^{(0)}$ denote the centralized UAV point clouds (with the same geometric center) of the ground truth and reconstruction result, respectively; $\boldsymbol{q}$ and $\boldsymbol{q}_{\mathsf{est}}$ correspond to the true and the estimated positions of the UAV based on the geometric centroid of the point cloud 3D coordinates. The first two terms in \eqref{WD-compute} correspond to the MCD that evaluates the reconstruction performance of the point cloud itself, and the remaining term is used to evaluate the accuracy of the UAV position estimation. In addition, to estimate the UAV attitude, we develop a heuristic attitude estimation method motivated by the fact that the UAV rotor plane is parallel to its fuselage plane. Specifically, we denote the partial points on the UAV rotor plane by the point set $\mathcal{S}_{\text{rotor}}$, and define the normalized normal vector of the UAV fuselage plane as $\boldsymbol{\varpi}\triangleq[\varpi_{\rm x},\varpi_{\rm y},\varpi_{\rm z}]^{\mathsf{T}}$. Then, we perform $J$ samplings by randomly selecting three points $\boldsymbol{p}_{a;\jmath}, \boldsymbol{p}_{b;\jmath}, \boldsymbol{p}_{c;\jmath}\in \mathcal{S}_{\text{rotor}}$ and the estimated normal vector $\boldsymbol{\varpi}_{\text{est}}$ can be obtained as:
\begin{align}
    \label{normal vector}
    \boldsymbol{\varpi}_{\text{est}} = \frac{1}{J}\sum^{J}_{\jmath=1}\boldsymbol{\varpi}_{\jmath} = \frac{1}{J}\sum^{J}_{\jmath=1}\frac{(\boldsymbol{p}_{b;\jmath}-\boldsymbol{p}_{a;\jmath})\times(\boldsymbol{p}_{c;\jmath}-\boldsymbol{p}_{a;\jmath})}{\|(\boldsymbol{p}_{b;\jmath}-\boldsymbol{p}_{a;\jmath})\times(\boldsymbol{p}_{c;\jmath}-\boldsymbol{p}_{a;\jmath})\|}.
\end{align}
Based on the cosine similarity, we consider a mean directional error (MDE) to evaluate the performance of UAV attitude (normal vector) estimation, which is defined as $\text{MDE} = \mathbb{E}[1-(\boldsymbol{\varpi})^{\mathsf{T}}\boldsymbol{\varpi}_{\text{est}}]$. All inference processes are performed on the Nvidia RTX 3090 platform, and the average inference time for each point cloud reconstruction is approximately 0.17 seconds when $M=1000$.

For performance comparison, we consider the following two benchmark schemes: \textbf{1)} A simplified approach that removes the position and SNR embedding (denoted as \textbf{NoEBD}); \textbf{2)}
A simplified approach that directly uses the estimated channel as a conditional prior for the diffusion model (denoted as \textbf{Direc-Channel}).  In addition, to compare the performance of UAV position estimation, the mean positioning error (MPE), i.e., $\text{MPE} = \mathbb{E}[\|\boldsymbol{q}-\boldsymbol{q}_{\mathsf{est}}\|]$, is used as an evaluation metric.

Figs. \ref{GT1} and \ref{Proposed_Noncenter} illustrate the EP point cloud reconstruction results of the proposed approach within the predicted flight region. It can be seen that, due to the estimation error in UAV position, the UAV may lie anywhere within the predetermined region. The proposed approach reconstructs the UAV EP point cloud with high fidelity, capturing its relative 3D position, attitude, and shape information within the reconstruction region. In this manner, the UAV position can be estimated by adding the reconstructed offset to the predicted center. This validates the effectiveness of the proposed approach for precise UAV positioning. The remaining point cloud plots in Fig. \ref{PC} present the centralized view of the point clouds reconstructed based on different approaches. Clearly, the proposed approach yields the EP point cloud substantially closer to the ground truth than the benchmark schemes, which show the lower-quality reconstruction with significant contamination by noisy points. This demonstrates that the encoder with the positional and SNR embeddings enhances the network's ability to accurately extract the intrinsic features of UAVs from the channel.

\begin{figure}[t]
    \centering
\includegraphics[width=.8\linewidth]{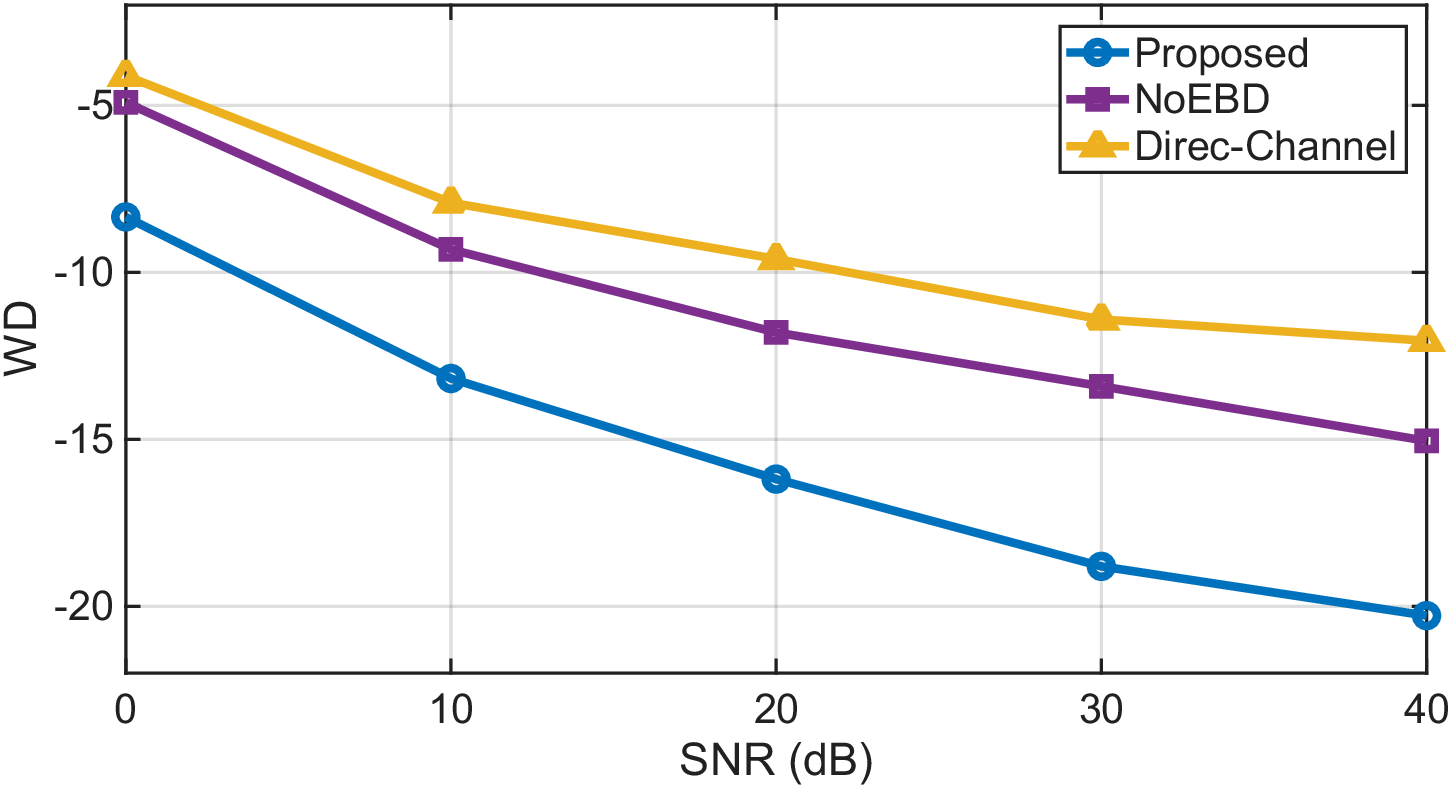}
    \caption{The WD comparison under varying SNR conditions.}
    \label{WD}
    \vspace{-0.3 cm}
\end{figure}

\begin{figure}[t]
    \centering
\includegraphics[width=.8\linewidth]{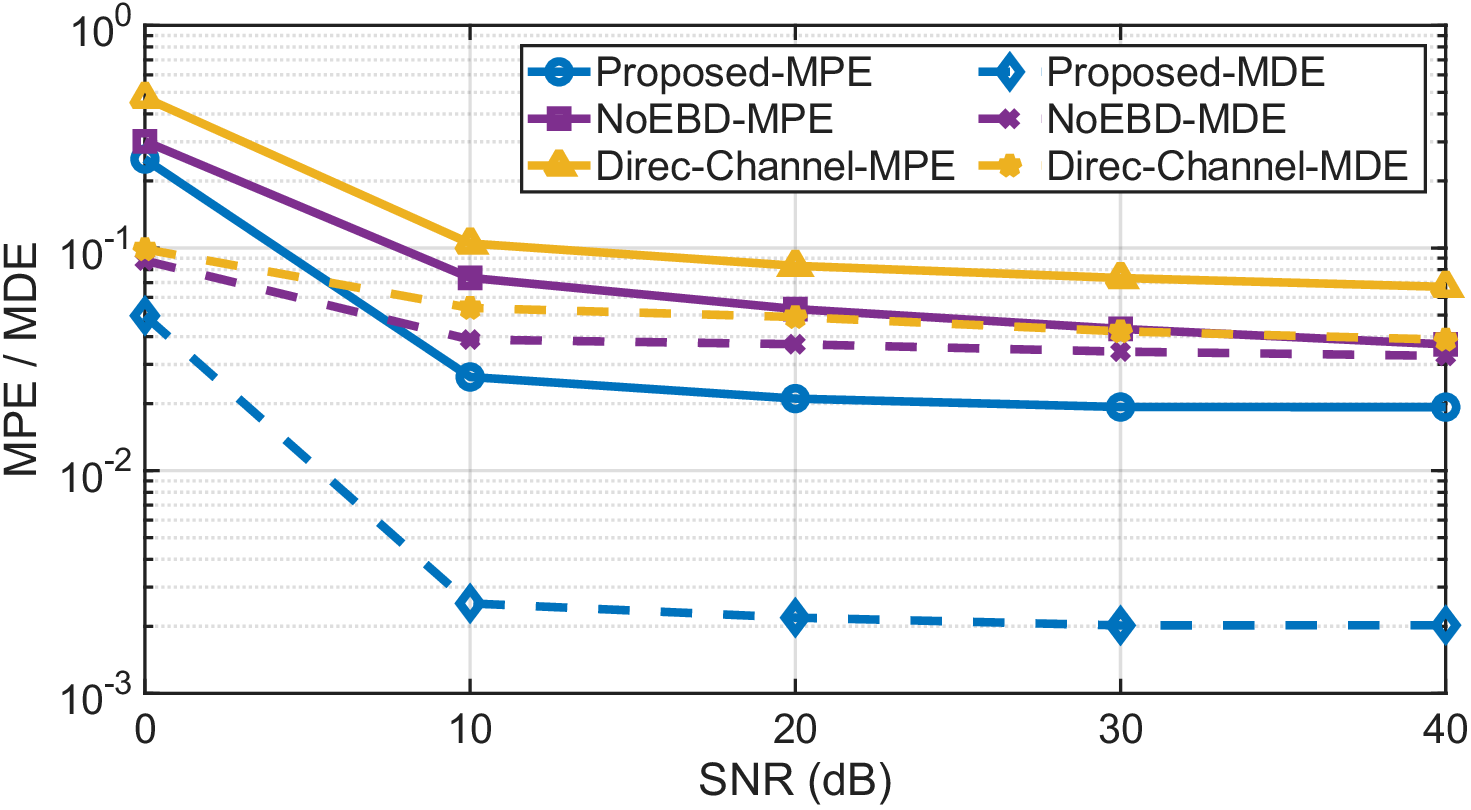}
    \caption{The MPE/MDE comparisons under varying SNR conditions.}
    \label{PDE}
    \vspace{-0.3 cm}
\end{figure}

The above conclusion is further supported by the quantitative results in Fig. \ref{WD} evaluated in terms of WD. Our proposed approach consistently achieves lower WD values across the considered SNR regimes, which means a higher fidelity point cloud imaging quality and positioning accuracy. Fig. \ref{PDE} compares the MPE and MDE of the considered approaches under different SNR conditions. It can be seen that when the SNR condition is slightly improved (i.e., $\text{SNR}\ge 10$ dB), the performances of all schemes are significantly improved. Our approach can achieve more precise UAV localization (close to centimeter level) within a given potential region compared to the benchmark schemes. Furthermore, due to the high-fidelity point cloud reconstruction capability, our proposed approach also significantly outperforms the benchmark scheme in UAV attitude estimation.

\section{Conclusion} 
Based on electromagnetic scattering modeling, we leveraged the estimated sensing channel and developed a conditional diffusion-based approach to achieve UAV position and attitude sensing via EP point cloud imaging within a potential flight region. The multiplicative position embedding and the SNR embedding are utilized to assist the MLP-based encoder in extracting the stable UAV features under varying UAV locations and channel conditions. Conditioned on the extracted features, a diffusion model is employed to reconstruct the UAV EP point cloud. Numerical results demonstrate that the proposed approach presents higher-fidelity point cloud imaging than the competing schemes, thereby enabling more accurate characterization of the UAV attitude and shape, and also the precise UAV positioning.
\ifCLASSOPTIONcaptionsoff
  \newpage
\fi

\bibliographystyle{IEEEtran} 
\bibliography{my_reference}

\end{document}